\documentclass[twocolumn,showpacs,preprintnumbers,amsmath,amssymb,floatfix]{revtex4}

\usepackage{graphicx}
\usepackage{dcolumn}
\usepackage{bm}
\usepackage{color}

\newcommand{\beq}{\begin{equation}}
\newcommand{\eeq}{\end{equation}}
\newcommand{\bey}{\begin{eqnarray}}
\newcommand{\eey}{\end{eqnarray}}

\begin{document}


\title{Dark energy stars: Stable configurations}

\author{Piyali Bhar}
 \email{piyalibhar90@gmail.com}
\affiliation {Department of Mathematics, Government General Degree College, Singur, Hooghly 712 409, West Bengal,
India}
\author{Tuhina Manna}
 \email{tuhinamanna03@gmail.com}
\affiliation {Department of Commerce (Evening), St. Xaviers College, 30 Mother Teresa Sarani,
Kolkata 700016, West Bengal, India}

\author{Farook Rahaman }
 \email{rahaman@associates.iucaa.in}
\affiliation {Department of Mathematics, Jadavpur University, Kolkata-700032, India}

\author{Ayan Banerjee}
 \email{ ayan_7575@yahoo.co.in}
\affiliation {Department of Mathematics, Jadavpur University, Kolkata-700032, India}

\date{\today}

\begin{abstract}
In present paper a spherically symmetric stellar configuration has been analyzed by assuming the matter distribution of the stellar configuration is anisotropic in nature and compared with the realistic objects, namely, the low mass X-ray binaries (LMXBs) and X-ray pulsars. The analytic solution has been obtained by utilizing the dark energy equation of state for the interior solution corresponding to the Schwarzschild exterior vacuum solution at the junction interface. Several physical properties like energy conditions, stability, mass-radius ratio, and surface redshift are described through mathematical calculations as well as graphical plots. It is found that obtained mass-radius ration of the compact stars candidates like 4U 1820-30, PSR J 1614-2230, Vela X-1 and Cen X- 3are very much consistent with the observed data by Gangopadhyay et al. (Mon. Not. R. Astron. Soc. 431, 3216 (2013)). So our proposed model would be useful in the investigation of the possible clustering of dark energy.
\end{abstract}


\maketitle

\section{Introduction}
In 2005 physicist George Chaplin proposed that, gravitational
collapse of objects with masses greater than a few solar masses
should lead to the formation of a compact object called {\em dark
energy star} whose surface corresponds to a quantum critical
surface for space-time, and whose interior differs from ordinary
space-time only in having a much larger vacuum energy
\cite{chapline1}. He claimed that the current picture of
gravitational collapse as explained in terms of event horizon is
physically inconsistent since it conflicts with quantum mechanics
\cite{chapline2}. The theory states that infalling matter gets
converted to dark energy as it falls through the event horizon
causing the space inside the event horizon to have a high value
of cosmological constant and hence a high negative pressure to
exert against gravity. This high negative pressure may counter
the mass the star gains thus avoiding a singularity. Thus there
is a sort of phase transition occurring in the phase of space at
the event horizon \cite{chapline2,stanley}. In fact  Dark
energy star may account for high energy cosmic ray sources and
positron sources since infalling masses decays into lighter
masses at the event horizon accelerating proton decay. This
proposal may provide a new perspective on spectacular
astrophysical phenomena including supernovae explosions, gamma
ray bursts, positron emission, and dark matter.\par

There is hardly any other topic in general relativity which is as extensively
researched as black holes. Since Oppenheimer and Snyder
\cite{Oppenheimer} first attempt at explaining the state of
gravitational collapse of a black hole, there have been various
observational data favouring the existence of event horizon, but
nonesoever proving it \cite{abramowicz}. This in turn has lead
to the developement of a plura of fascinating alternative ideas.
One such popular model is that of a gravastar or gravitationally
vacuum star as proposed by Mazur and Mottola \cite{mazur}. Its a
compact object  having an interior de Sitter condensate defined
by the equation of state $p=-\rho$ , matched to a shell of finite
thickness with an equation of state $p=\rho$ , which is again
matched to an exterior Schwarzschild vacuum solution. This star
has no singularity at the origin and no event horizon since the
thick shell surface with a radius slightly greater than the
Schwarzschild radius, replaces both the de Sitter and the
Schwarzschild horizons. The charged and uncharged model of gravastar was obtained
in our earlier works \cite{Horvat}. A modified version of the gravastar was
studied by Cattoen {\em et al.} \cite{cattoen} using a continuous
pressure profile without thin shells. It strongly suggested an
anisotropic pressure model. Later an alternative model termed as
Born-Infeld phantom gravastar was constructed replacing the
de-Sitter spacetime by an interior spacetime governed by the
Chaplygin gas equation of state \cite{born}. In this paper the authors have
investigated a generalized  case in which the equation of state
is governed by the equation of state $p=\omega \rho$ matched to
an exterior vacuum solution. The proposed stellar model consists
of five zones : an interior core, a thin shell between the core
and the interior spacetime, the interior spacetime and a thin
shell which acts as a junction interface between the interior and
the exterior Schwarzschild spacetime. This kind of compact object
which generalizes a gravatus model by using equation of state
~$p=\omega \rho$ with $\omega<-1/3$ ~is called a dark energy
gravastar or simply a dark energy star , as referred by Chapline.
This terminology is motivated by the fact that dark energy is
still unknown component of our Universe has relativistic negative
pressure. The 1998 observations of Type la Supernova
\cite{supernova} confirms that dark energy is responsible for the
phase of cosmic acceleration. Numerous other recent observations
of Cosmic Microwave Background(CMB) anisotropies and Large Scale
Structure(LSS) \cite{CMB} reconfirm this characteristics of small
redshift evolution of our Universe. The simplest explanation of
dark energy is the cosmological constant $\Lambda$ which is
usually interpreted physically as a vacuum energy, with
$p=-\rho$. Another possible way to explain the dark energy is by
invoking an equation of state, $p=\omega\rho$ with $\omega<0$ ,
where $p$ is the spatially homogeneous pressure and $\rho$ the
energy density of the dark energy, instead of the constant vacuum
energy density. The models of dark energy with $\omega>-1$ mainly
comprises of quintessence , k-essence and chaplygin gas among
others.

This idea of dark energy star theorizes that the surface of a compact object is a
quantum critical shell with some thickness \textit{{$z$}} \cite{chapline3}. When ordinary elementary
particles which have energy beyond $Q_0= 100 MeV\sqrt{\frac{M_0}{M}}$, where $M_0$ is the solar mass,
enter this quantum critical region they decay into substituent products and radiation
that gets directed backwards from the surface of dark energy star perpendicular to
the critical surface. However for particles with energy $<Q_0$ will pass through the critical
surface and follow diverging geodesics in the interior of a dark energy star. Now compact
objects at the centre of galaxies contains quarks and gluons inside nucleons whose energies exceed
this value \cite{chapline4}. According to Georgi-Glashow grand unified model nucleons
can decay by a process in which a quark decays into a positron and two antiquarks. Interestingly an
excess of positrons have been detected in the centre of galaxies which may be the best
evidence for dark energy stars. In addition to that, primordial dark energy stars  can form out of
fluctuations of spacetime analogous to a quantum critical instability \cite{chapline2}.
This in turn can help us understand dark matter. Thus motivated dark energy star have been
investigated in many notable work. In a pioneering paper Lobo \cite{lobo} has given a
model of a stable dark energy star by assuming two spatial types of mass function: one is of constant
energy density and the other mass function is a Tolman\& Whitker mass.
All the features of the dark energy star have been discussed and the system is found to
be stable under a small linear perturbation. In our previous work a new model of dark energy star
consisting of five zones, namely, the solid core of constant energy density, the thin
shell between core and interior, an inhomogeneous interior region with anisotropic pressures, a thin shell,
and the exterior vacuum region is obtained \cite{pb15}. Yadav {\em et al.} \cite{y} have
given a dark energy model with a variable equation of state parameter. Inspired by the
prior mentioned works of Cattoen {\em et al.} \cite{cattoen} and Ruderman(1972) \cite{ruderman}
we have taken the pressure inside the fluid sphere in our model to be anisotropic. For an
anisotropy distribution, the pressure inside the fluid sphere is decomposed into two orthogonal
components: radial pressure $p_r$ and transverse pressure $p_t$ where obviously $p_r\neq p_t$.
Studies on X-ray pulsars, Her-x-1, X-ray buster 4U 1820-30, millisecond pulsar SAXJ1804.4-3658 etc.
suggests that nuclear matter tends to become anisotropic in nature at very high
densities ( $10^{15}gm/cc$.) \cite{herrerasantos,bowers}. Anisotropy may occur due to any
of the following reasons : existence of solid core, a type 3A superfluid \cite{kw}, phase
transition \cite{sokolov}, pion condensation \cite{sawyer}, rotation, magnetic field, mixture
of two fluid, existence of external field etc.\par

We organised the paper as follows : In Sec. \textbf{II.} deals with the basic field
equations and their solutions, then the junction conditions
have been discussed in Sec. \textbf{III}.  In Sec. \textbf{IV.} the physical properties
with the stability conditions have been studied. Finally, in Sec. \textbf{V.}
we discuss some specific comments regarding the results obtained in the study.

\section{Interior space-time and the Field equations}

We consider  a static spherically symmetric space-time given by
the following line element:
\begin{equation}
ds^{2}= -exp\left[{-2\int_r^{\infty}g(\tilde{r})d\tilde{r}}\right]dt^{2} +\frac{dr^{2}} {1-\frac{2m}{r}}+
r^2 d\Omega^2 ,
\end{equation}
where $r$ denotes the radial coordinate. Here $g(r)$ denote an arbitrary function of the radial coordinate representing the locally measured gravitational acceleration, which according to general convention is assumed to be positive in case of an inward directed gravitational
attraction, and negative in the case an outward directed gravitational repulsion. Also the function $m(r)$ denotes the gravitational mass contained within a sphere of radius $r$.

We proceed now to describe the matter distribution of the spherical body which is anisotropic in nature,
and the energy-momentum tensor is characterized by the following relationships
\begin{equation}
T_{\mu\nu}=(\rho+p_t)U_{\mu}U_{\mu}+p_t g_{\mu\nu}+(p_r-p_t)\chi_{\mu}\chi_{\nu},
\end{equation}
where $U_{\mu}$ is the four-velocity and $\chi_{\mu}$ is the unit spacelike
vector in the radial direction. This interpretation
can be justified by invoking the static stress-energy tensor
$T^{\mu} _\nu$ = diag[ $\rho$, $p_r$, $p_t$, $p_t$]. Now, using the energy-momentum tensor for the metric given
in Eq. (1), the Einstein field equation provides the following relationships (in relativistic units with G = c = 1)
\begin{equation}
m'=4\pi r^{2}\rho,
\end{equation}
\begin{equation}
g=\frac{m+4\pi r^{3}p_r}{r(r-2m)},
\end{equation}
\begin{equation}
p_r'=-\frac{(\rho+p_r)(m+4\pi r^{3}p_r)}{r(r-2m)}+\frac{2}{r}(p_t-p_r),
\end{equation}
where $\rho(r)$ is the energy density, $p_r(r)$ and $p_t(r)$ are the radial and
tangential pressures respectively and `prime' denotes the derivative with respect
to the radial co-ordinate, $r$. Equation (5), corresponds to the Bianchi identity implies
that $\Delta_\mu T^{\mu\nu}=0$, may also obtained by using relativistic
Tolman-Oppenheimer-Volkov (TOV) equation for anisotropic pressure.

Thus we have three equations, namely, the field equations (3)-(5), with five unknown
functions of r, i.e., $\rho$(r), $p_r$(r), $p_t$(r), g(r) and m(r). Therefore, it is
extremely difficult to obtain an explicit solutions of Einstein field equations.
But as in aforesaid discussion, we are interested in more realistic situation
where the mass function is uniform and field distributions are simply obtained
for determining the physical features of a compact star. With this aim in mind,
let us assume that the mass density is increasing from the star center ( r = 0 )
to the surface ( r = R ) and having an equation of state $p_r=p_r(\rho)$.

At this point we introduce the mass function
\begin{equation}
m(r)=\frac{ar^{3}}{2}\left(\frac{2+ar^{2}}{(1+ar^{2})^{2}}\right),
\end{equation}
along with the dark  energy equation of state in the particular form
\begin{equation}
p_r = \omega \rho.
\end{equation}
For the choice of mass function gives a monotonic decreasing matter density as used
earlier by Singh et al.\cite{singh}, for modelling a compact star of embedding class I.
It will be suitable to express the physical variables in terms of metric function,
and our anisotropic fluid configuration to be physically acceptable.
Now, taking into account the Eqs. (3) and (4) and using the
dark energy equation of state, we obtain
\begin{eqnarray}\label{gr}
g(r)=\frac{ar}{2(1+ar^2)}\left[2(1+3\omega)+(1+\omega)(3ar^2+a^2r^4)\right],
\end{eqnarray}
To study the nature of dark energy for local astrophysical manifestation
has attracted a lot of interest and for excellent reviews on this topic see Ref. \cite{Stoytcho}.
The existence of dark energy star makes us expect that it is a generalization of
the gravastar picture with an interior solution governed by the dark energy EOS.Such objects have received considerable attention in astrophysics though some steps in this direction have already emerged.

Solving the above differential Eqs. (3-5), using dark energy equation of state $p_r = \omega \rho$,
we obtain the physical parameters for this model are then given by :
\begin{eqnarray}
\rho&=&\frac{a}{8\pi}\left[\frac{6+3ar^2+a^2r^4}{(1+ar^2)^3}\right],\\
p_r&=&\frac{\omega a}{8\pi}\left[\frac{6+3ar^2+a^2r^4}{(1+ar^2)^3}\right],\\
p_t&=&\frac{a}{32\pi(1+ar^2)^4}\left[12ar^2+24a^2r^4+17a^3r^6+6a^4r^8\right.\nonumber\\
&&\left. +a^5r^{10}+\omega (24+24ar^2+60a^2r^4+38a^3r^6\right.\nonumber\\
&&\left.+12a^4r^8 +2a^5r^{10})+\omega^2(36ar^2+36a^2r^4+21a^3r^6\right.\nonumber\\
&&\left. +6a^4r^8+a^5r^{10})\right],
\end{eqnarray}
and the anisotropic factor $\Delta$ is given by
\begin{eqnarray}
\Delta&=&p_t-p_r=\frac{a}{32\pi(1+ar^2)^4}\left[12ar^2+24a^2r^4+17a^3r^6\right.\nonumber\\
&&\left.+6a^4r^8+a^5r^{10}+\omega (-12ar^2+44a^2r^4+34a^3r^6\right.\nonumber\\
&&\left. +12a^4r^8+2a^5r^{10})+\omega^2(36ar^2+36a^2r^4+21a^3r^6\right.\nonumber\\
&&\left. +6a^4r^8+a^5r^{10})\right].
\end{eqnarray}
The anisotropy factor measures the pressure anisotropy of the fluid comprising
the dark energy star. An isotropic pressure dark energy star corresponds to $\Delta=0$.
Here $\Delta/r$ represents a force arising due to the anisotropic nature of the stellar model.
The anisotropy will be repulsive or directed outwards if $p_t > p_r$, and attractive or directed inward
when $p_t < p_r$. From the above set of solutions we observe that
the central density, $\rho_0$= 6a/8$\pi$,  being a non-zero constant quantity, satisfy the
regularity conditions and finite character at the origin.

We shall now analyze our model and verify that the energy density is
positive and finite at all points in the interior of the star and some
physical criterion which are necessary for a stable interior solution. Therefore, it is
necessary to impose the restrictions on the constants appearing in the metric functions,
so that all criteria for physical acceptability are satisfied and well behaved at all the inner
points of strange stars. So, taking derivatives of Eqs. (9), (10) and (12) with respect to
the radial coordinate, we have
\begin{eqnarray}
\frac{d\rho}{dr}&=&-\frac{a^2r}{4\pi}\left[\frac{15+4ar^2+a^2r^4}{(1+ar^2)^4}\right],\\
\frac{dp_r}{dr}&=&-\frac{\omega a^2r}{4\pi}\left[\frac{15+4ar^2+a^2r^4}{(1+ar^2)^4}\right], \\
\frac{d\Delta}{dr}&=&\frac{a}{32\pi(1+ar^2)^5}\left[24ar+24a^2r^3+6a^3r^5+14a^4r^7 \right.\nonumber\\
&&\left. +10a^5r^9+2a^6r^{11}+\omega(-24ar+248a^2r^3+28a^3r^5\right.\nonumber\\
&&\left. +28a^4r^7+20a^5r^9+4a^6r^{11})+\omega^2(72ar-72a^2r^3\right.\nonumber\\
&&\left. -82a^3r^5+6a^4r^7+10a^5r^9+2a^6r^{11})\right].\\
\end{eqnarray}
Here, we observe that $\frac{d\rho}{dr} =0$ and $\frac{dp_r}{dr} = 0$ at the origin, which
clearly indicate that at the centre of the star, density is maximum
and it decreases radially outward. We have also verified that the
radial pressure is maximum at the centre and it decreases towards the boundary.
Thus, the energy density is continuous and well behaved in the stellar interior has
been shown in Fig. \textbf{1}. One readily verifies from Figs. \textbf{2} and \textbf{3},
that  g(r) $>$ 0, within the range of $-1/3<\omega<0 $, indicating an inward
gravitational attraction, and  within the range of $ -1/3<\omega<-1$, g(r) $<$ 0, shows an outward
gravitational repulsion. Since, we are dealing with dark energy star, thus it is necessary that
the interior solution must be repulsive in nature, so that the region where g(r) $>$ 0
associate with inward gravitational attraction is necessarily excluded.

The entire analysis has been performed with a set of astrophysical objects in
connection to direct comparison of some strange/compact star candidates
like X-ray pulsar Cen-X3, X-ray burster 4U 1538-52, and Xray sources 4U 1538-52.
In connection with that we extensively studied the physical features of compact star
Vela X - 1.

\section{Junction Condition}
Now, we analyse the junction conditions, where the
space-time is matched to an exterior schwarzschild
vacuum solution with p = $\rho$ = 0 at the junction interface $\Sigma$,
with junction radius R. The line element of the Schwarzschild exterior spacetime is given by:
\begin{equation}
ds^{2}=-\left(1-\frac{2M}{r}\right)dt^{2}+\frac{dr^{2}}
{1-\frac{2M}{r}}+r^{2}(d\theta^{2}+\sin^{2}\theta d\phi^{2}).
\end{equation}
We know that the event horizon for this spacetime occurs at $r_h$ = 2M. Since our proposed model of dark energy star does not possess any event horizon we choose the value of
junction radius R $> r_h$ i.e., the junction radius lies outside 2M.
Now using the standard Darmois-Israel formalism \cite{Israel}, the intrinsic stress-energy tensor
$S_{ij}$, at the junction surface $\Sigma$ is defined through the Lanczos equation as
\begin{equation}
S^{i}_{j} = -\frac{1}{8\pi}\left(\kappa^i_j-\delta^i_j\kappa^m_m\right),
\end{equation}
where the quantity $k_{ij}$ represents the discontinuity in the
extrinsic curvature $K_{ij}$ and the discontinuity is given by
$k_{ij} = K^{+}_{ij}-K^{-}_{ij}$. Each of the extrinsic curvatures and using the second fundamental
form the extrinsic curvature can always be written as
\begin{equation}
K^{\pm}_{ij} = -\eta_{\nu}\left(\frac{\partial^2 x^{\nu}}{\partial\xi^i \partial\xi^j}+\Gamma^{\nu\pm}_{\alpha\beta}\frac{\partial x^{\alpha}}{\partial\xi^{i}}\frac{\partial x^{\beta}}{\partial\xi^{j}}\right),
\end{equation}
where $\eta_{\nu}$ are the units normal at the junction $\Sigma$.
The symbol $`\pm'$ corresponding to the interior and exterior spacetime, and $\xi^i$ represents
the intrinsic co-ordinates on $\Sigma$. As a result we can write down the non-trivial
components of the extrinsic curvature by using the metrics (1) and (16), are given by
\begin{equation}
K^{\tau+}_\tau=\frac{\frac{M}{R^2}+\ddot{R}}{\sqrt{1-\frac{2M}{R}+\dot{R}^2}},
\end{equation}
\begin{eqnarray}
K^{\tau-}_{\tau}&=&\frac{1}{\sqrt{\dot{R}^2+(1+aR^2)^{-2}}}\left[\ddot{R}\right.\nonumber\\
&&\left.+\frac{aR\left[(2+6\omega)+(1+\omega)(3aR^2+a^2R^4)\right]}
{2(1+aR^2)^3}\right.\nonumber\\
&&\left.-\frac{(1+\omega)\dot{R}^2aR(6+3aR^2+a^2R^4)}{2(1+aR^2)(1+4aR^2+2a^2R^4)}\right],
\end{eqnarray}
\begin{equation}
K^{\theta+}_\theta=\frac{1}{R}\sqrt{1-\frac{2M}{R}+\dot{R}^2},
\end{equation}
\begin{equation}
K^{\theta-}_{\theta}=\frac{1}{R}\sqrt{\dot{R}^2+(1+aR^2)^{-2}},
\end{equation}
where the overdot denotes derivative with respect to proper time
$\tau$. Thus, the Lanczos equations \cite{lobo} allow us to write the surface stresses,
as follows:
\begin{equation}
\sigma=-\frac{1}{4\pi~R}\left[\sqrt{1-\frac{2M}{R}+\dot{R}^2}-\sqrt{\frac{1}{(1+aR^2)^2}+\dot{R}^2}\right],
\end{equation}
and
\begin{eqnarray}\label{eq:P}
\mathcal{P}&=&\frac{1}{8\pi~R}\left[\frac{1-\frac{M}{R}+\dot{R}^2+R\ddot{R}}{\sqrt{1-\frac{2M}{R}+\dot{R}^2}}\right.\nonumber\\
&&\left.-\frac{(1+\omega)\dot{R}^2aR^2(6+3aR^2+a^2R^4)}{2(1+aR^2)(1+4aR^2+2a^2R^4)}\right.\nonumber\\
&&\left.+\frac{2+(4+6\omega)aR^2+(1+\omega)(3a^2R^4+a^3R^6)}{2(1+aR^2)^3}\right.\nonumber\\
&&\left.+\dot{R}^2+R\ddot{R}{\sqrt{(1+aR^2)^{-2}+\dot{R}^2}}\right].
\end{eqnarray}

 The corresponding interpretation of $\sigma$ as the energy-density, while $\mathcal{P}$
as the tangential surface pressure connected with the work
done by the internal forces on the junction surface.

It is then necessary to use the conservation identity given by
$S^i_j|_i=\left[T_{\mu\nu}e^\mu_{(j)}n^\nu\right]^{+}_-$,
where $n^\mu$ is the unit normal 4-vector to $\Sigma$ and $e^\mu_{(i)}$ are
components of the holonomic basis vectors tangent to $\Sigma$, where we are defining
$[X]^+_-$  the discontinuity across the junction surface i.e., $[X]^+_-=X^+|_\Sigma-X^-|_\Sigma $.

From the above expression of conservation we can write
$ S^i_\tau|_i=-[\dot{\sigma}+2\dot{R}(\sigma+\mathcal{P})/R]$, where
the momentum flux term in the right hand side of the conservation identity is given by
\begin{equation}
\left[T_{\mu\nu}e^\mu_{(j)}n^\nu\right]^+_-
 =-\frac{(\rho+p_r)\dot{R}\sqrt{1-2m/R+\dot{R}^2}}{1-2m/R}.
\end{equation}
where $\rho$ and $p_r$ may be deduced from Eqs. (3-4), respectively,
evaluated at the junction radius, R. Further it is easy to see from
two Eqs. (23) and (24) that the energy conservation equation is fulfilled
and can be recast as \cite{lobocrawford}:

\begin{equation}\label{eq:diffsigma}
\sigma^\prime=-\frac{2}{R}(\sigma+\mathcal{P})+\Xi,
\end{equation}
where we are defining the prime operation as the derivative
with respect to $R$ and $\Xi$ is given by
\begin{equation}
\Xi=-\frac{1}{4\pi~R}\frac{(1+\omega)aR(6+3aR^2+a^2R^4)}{2(1+ar^2)(1+4aR^2+2a^2R^4)}\sqrt{(1+aR^2)^{-2}+\dot{R}^2}.
\end{equation}
It is instructive to analyze the flux term [Eq. (27)] is zero when
$w = -1$, which reduces to the analysis, see Ref. \cite{Visser}. We now proceed with the
surface mass of the thin shell is given by
\begin{equation}
m_s=4\pi~R^2\sigma.
\end{equation}
When one substitutes Eq. (23) into Eq. (28), one can get
\begin{equation}
m_s=-R\left[\sqrt{1-\frac{2M}{R}+\dot{R}^2}-\sqrt{\dot{R}^2+(1+aR^2)^{-2}}\right],
\end{equation}
It is interesting to note that rearranging the above expression and evaluating at $R = R_0$, lead us to interpret M as the total mass of the dark energy star, as given below,
\begin{equation}
M=\frac{aR_0^3}{2}\frac{2+aR_0^2}{(1+aR_0^2)^2}+m_s(R_0)\left(\frac{1}{(1+aR_0^2)^2}-\frac{m_s(R_0)}{2R_0}\right).
\end{equation}

Now, differentiating twice the expression in (28), and taking into account the radial
derivative of $\sigma^\prime$, we start by rewriting Eq. (26) in a convincing form
\begin{equation}
\left(\frac{m_s}{2R}\right)^{\prime\prime}=\Upsilon-4\pi~\sigma^\prime\eta,
\end{equation}
where the unknown parameters are given by
\begin{equation}
\eta=\frac{\mathcal{P}^\prime}{\sigma^\prime},~~~~\Upsilon=\frac{4\pi}{R}(\sigma+\mathcal{P})+2\pi~R\Xi^\prime.
\end{equation}
The expression found above will play an important role in stability analysis of static solutions.
Here the parameter $\eta$ is used to determine the stability of the
system. There are ranges of the parameters $\sqrt{\eta}$ for which the stability regions
looks more convincing when $0 < \eta \leq 1$, so $\eta$ can be interpreted as the velocity of
sound on the shell.

In order to gain more impression we analyze our system by evolution identity and
given by: $\left[T_{\mu\nu}n^\mu_{(j)}n^\nu\right]^+_-= \bar{K}^i_jS^j_i$,
where $\bar{K}^i_j=(K^{i+}_j+K^{i-}_j)/2$. The above discussion leads to the
evolution identity can be written as follows:
\begin{eqnarray}
p_r+\frac{(\rho+p_r)\dot{R}^2}{1-2m/R}=\nonumber\\
-\frac{1}{R}\left[\sqrt{1-\frac{2M}{R}+\dot{R}^2}
+\sqrt{\frac{1}{(1+aR^2)^2}+\dot{R}^2}\right]\mathcal{P}\nonumber\\
+\frac{1}{2}\left[\frac{M/R^2+\ddot{R}}{\sqrt{1-\frac{2M}{R}+\dot{R}^2}}\right.
\left.+\frac{1}{\sqrt{(1+aR^2)^{-2}+\dot{R}^2}}\times\nonumber\right.\\
\left.\left(\frac{aR[2+6\omega+(1+\omega)(3aR^2+a^2R^4)]}{2(1+aR^2)^3}\right.\right.\nonumber\\
\left.\left.+\ddot{R}+\frac{aR\dot{R}^2(1+\omega)(6+3aR^2+a^2R^4)}{2(1+aR^2)(1+4aR^2+2a^2R^4)}\right)\right]\sigma
\end{eqnarray}
For the static solution at $R = R_0$  with $\dot{R}=\ddot{R}=0$, we can obtain the from of
above equation the radial pressure in terms of the surface stress as :
\begin{eqnarray}
p_r(R_0)&=&-\frac{1}{R_0}\left(\sqrt{1-\frac{2M}{R_0}}+\frac{1}{\sqrt{1+aR_0^2}}\right)\mathcal{P}\nonumber\\
&&+\frac{1}{2R_0^2}\left(\frac{M}{\sqrt{1-\frac{2M}{R_0}}}+\frac{aR_0^3}{2(1+aR_0^2)^2}\left[(2+6\omega)\right.\right.\nonumber\\
&&\left.\left.+(1+\omega)(3aR_0^2+a^2R_0^4)\right]\right)\sigma.
\end{eqnarray}
We have seen that $\sigma < 0$ and the shell pressure acting from the interior is negating
$p_r < 0$, implying that the tension is in the radial direction.
Therefore, a positive tangential surface pressure $\mathcal{P}$ is required for a stable thin shell and
to prevent it from collapsing.

\section{ Physical properties and Comparative study of the physical parameters for
dark energy star model}
In order to obtain a feasible solution of a dark energy star, the exact solutions of Einstein's field
equations must satisfy some general physical requirements. Our goal is to explore the
physical features of the strange dark star and determine the constraints for
which the solutions are physically realistic. Next, we carry out a comparative study
with the most recent observational data of number of compact objects given by Gangopadhyay
et al. \citep{gan}. Following which we analyze in detail the desirable physical
properties of this star solution, in particular, free from physical and geometrical
singularities, energy density and pressure are finite character at the origin r = 0, satisfying
all the energy conditions at the stellar interior. We then proceed to solving
them mathematically and discuss the obtained solutions
with observational data, where we have considered the compact star Vela~ X - 1,
estimated mass M = $1.77 M_{\odot}$ and radius $9.56$.

\subsection{ \emph{Mass-radius relation and surface gravitational red shift}}
One can obtain the mass of the star within a radius r from Eq. (6).  It is
clear that the mass function $m(r)\rightarrow 0$ as $r\rightarrow 0$, which indicated
that mass function is regular at the center. The present paper highlights the maximum
allowable mass-radius ratio for a static spherically symmetric perfect fluid sphere falls
within the limit of $< 8/9$ in \cite{buchdahl} (in the unit c = G = 1).
Furthermore, considering the particular value of mass and radius for Vela X - 1,
we obtain the variation of mass function in the stellar interior, which
is depicted in Fig. \textbf{5}. Note that the mass function represents a monotonic
increasing function of `r' and $m(r) > 0$ when $0 < r < R$ i.e., within the radius of the star.
Similarly assuming the estimated masses and radii for several compact objects such as
X-ray pulsar Cen-X3, X-ray burster 4U 1538-52, and X-ray sources 4U 1538-52, we have performed
a comparative study of the values of the physical parameters which is shown in Table~I, and
are closely equal to the observed values of most of the stars. Once again, we compared our
solution of mass-radius relation for the different strange star, which do not cross the proposed
range by Buchdahl in \cite{buchdahl}.

The compactness of the strange stars are found under the assumptions
\begin{equation}
u(r)=\frac{m(r)}{r}=\frac{ar^2}{2}\frac{2+ar^2}{(1+ar^2)^2}.
\end{equation}
The compactness (u(r)) of the star is a monotonic increasing function of r and
the redshift function $z_s $, of the strange compact star can be defined through the equation
\begin{equation}
1+z_s=(1-2u)^{-\frac{1}{2}},
\end{equation}
where the surface redshift function $z_s$ is given by $z_s=ar^2$.
We have calculated the maximum surface redshift ($z_s$) for different strange stars
from our model which  shown in Table II. From the table it is clear that
maximum value of the surface redshift, $z_s < 1$. So our result is compatible
with the result obtained by \cite{hamity}.

\begin{figure}[htbp]
    \centering
        \includegraphics[scale=.3]{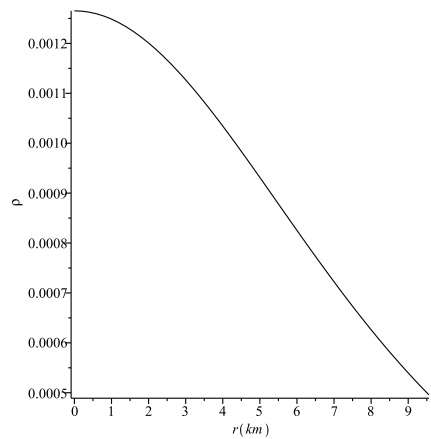}
       \caption{Density variation in GeV/$fm^3$ inside the star Vela~ X-1 with radial distance r in km. }
    \label{rho}
\end{figure}

\begin{figure}[htbp]
    \centering
        \includegraphics[scale=.4]{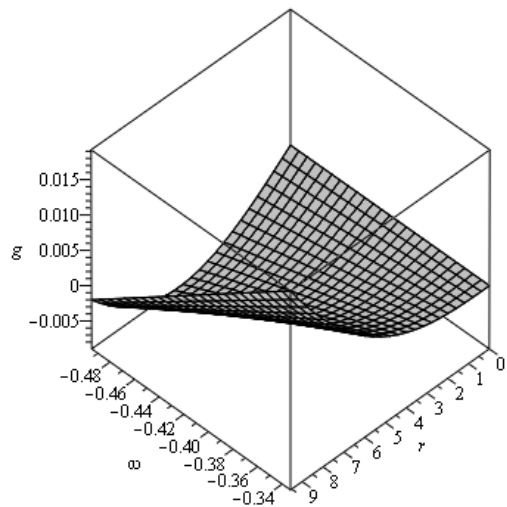}
       \caption{ The ``gravity profile",~$g(r)$,~ is plotted against $r$
when $-1/3<\omega<0$. Note that within the range of $-1/3<\omega<0 $, g(r) $>$ 0,
shows an inward gravitational attraction, so it is  necessarily excluded for our dark star model.}
    \label{gr2}
\end{figure}

\begin{figure}[htbp]
    \centering
        \includegraphics[scale=.4]{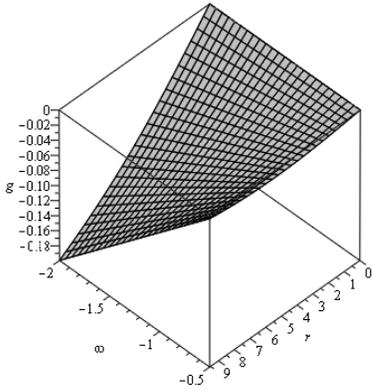}
       \caption{ The ``gravity profile",~$g(r)$,~ is plotted against $r$
when $-1/3<\omega<0$, for our estimated mass and radius. }
    \label{gr1}
\end{figure}

\begin{figure}[htbp]
\centering
\includegraphics[scale=.3]{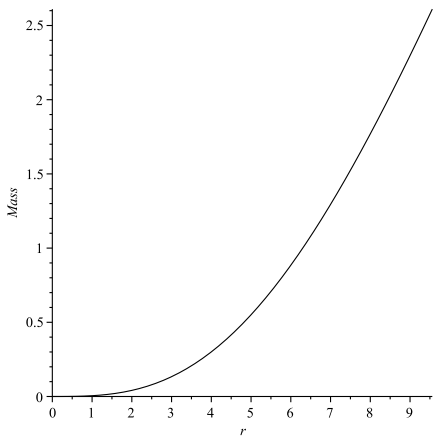}
\caption{ Mass-radius relationship of strange stars Vela~ X - 1, which is
characterised by the estimated mass M = $1.77 M_{\odot}$ and radius $9.56$.
The estimated masses of several compact objects, whose properties are given in Table I.}
\label{mass}
\end{figure}

\begin{figure}[htbp]
\centering
\includegraphics[scale=.4]{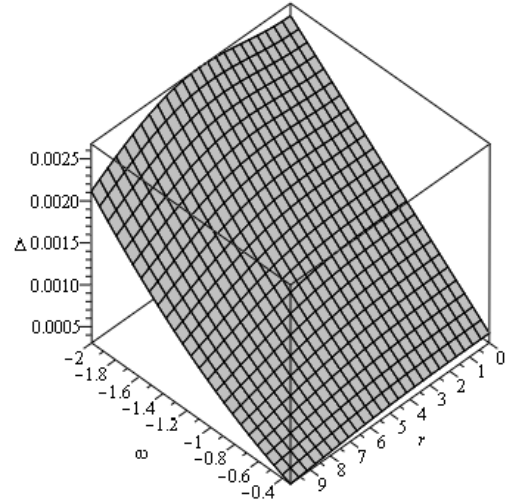}
\caption{ Anisotropy factor $\Delta$  at the interior of the star for $\omega<-1$.
We verify that $\Delta > 0$ in the phantom region when $\omega<-1$, of strange stars Vela X - 1
}
\label{delta}
\end{figure}

\begin{figure}[htbp]
\centering
 \includegraphics[scale=.3]{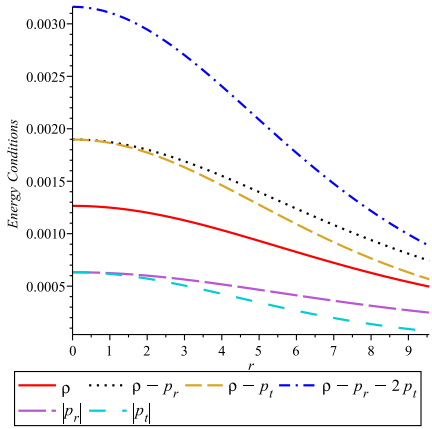}
 \caption{Energy Conditions inside the strange star `Vela~ X-1' is shown against r. From the figure it is clear that all the above mentioned energy conditions are satisfied by our model.}
 \label{ec}
\end{figure}

\begin{figure}[htbp]
    \centering
        \includegraphics[scale=.3]{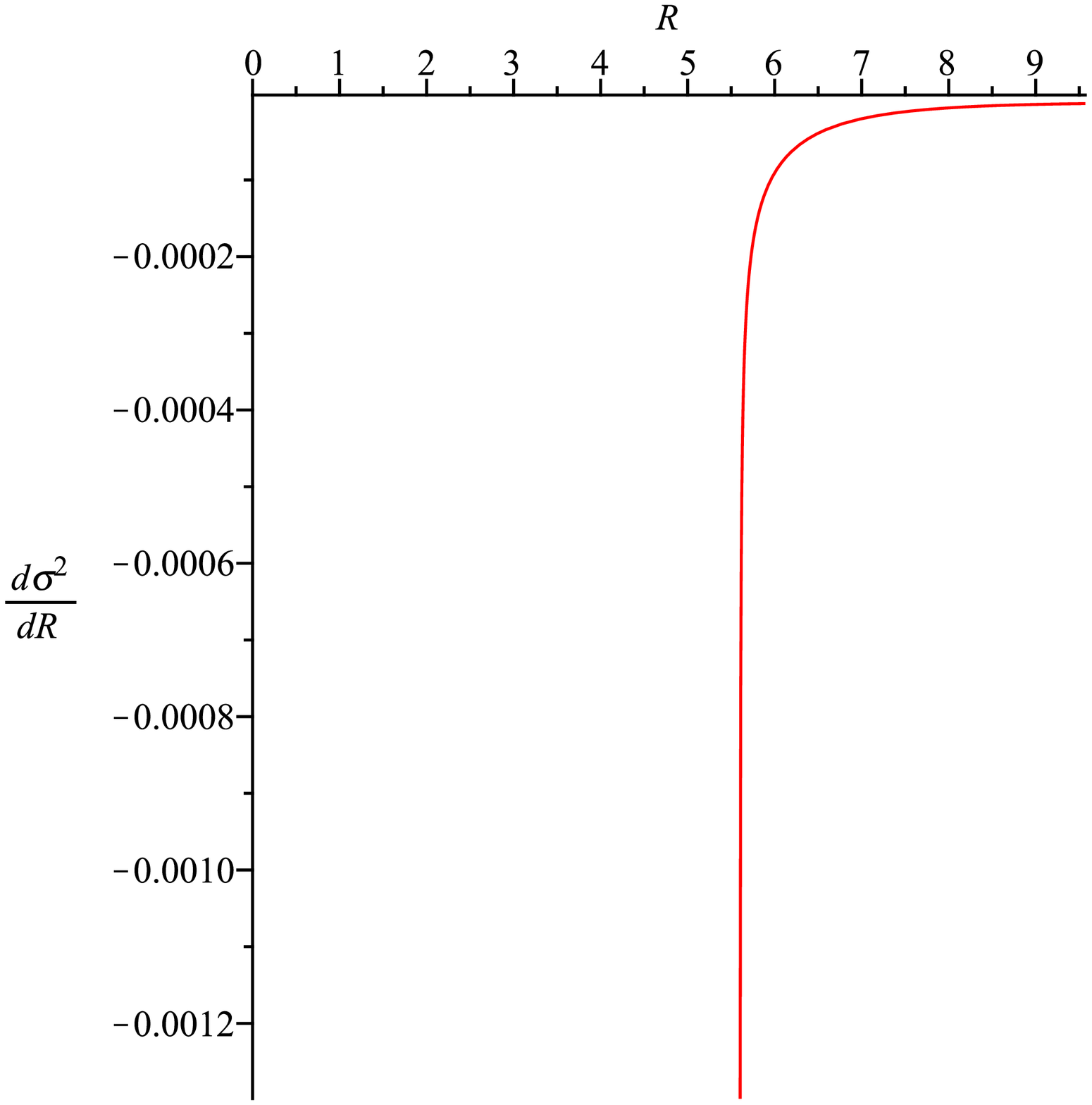}
       \caption{`Vela~ X-1' (radius R= 9.56): In these figure we have plotted the dimensionless
parameter L = $\frac{d\sigma^2}{dR}$. Therefore the stability region is constrained
by the inequality (52).}
    \label{dsigma}
\end{figure}

\subsection{\emph{Energy Condition}}
In this section we are going to verify whether our particular
model for dark energy star satisfies all the energy
conditions or not, namely, null energy condition (NEC), weak energy
condition (WEC), strong energy condition(SEC) and
dominant energy condition(DEC), at all points in the interior of a star
if the following inequalities hold simultaneously:
\begin{eqnarray}
\textbf{NEC:} \rho(r)-p_r \geq  0,\\
\textbf{WEC:} \rho(r)-p_r(r) \geq  0~~, ~~\rho \geq  0,\\
\textbf{SEC:} \rho(r)-p_r(r) \geq  0~~, ~~\rho-p_r(r)-2p_t(r) \geq  0,\\
\textbf{DEC:} \rho(r)> |p_r(r)| ~~, ~~\rho > |p_t|.
\end{eqnarray}
We show the above inequalities with the help of graphical representation.
In Fig. \ref{ec}, we have plotted the L.H.S of the above inequalities which
shows that our model satisfies these conditions for specific values of mass
and radius at the stellar interior when $\omega= -0.5$.

\begin{table*}
\caption{Values of the constants $a$, $b$, $H$ and $K$ for different compact star models.}
\begin{tabular*}{\textwidth}{@{\extracolsep{\fill}}llllllllll@{}}
\hline
Compact Star &  $a$ &$R~$(km) & $M/M_{\odot}$ &$R~$(km) & $M/M_{\odot}$&References \\
&&observed&observed&calculated&calculated\\
\hline
4U~1538-52&0.0037&$7.866\pm0.21$&$0.87\pm0.07$&7.87&0.90& Gangopadhyay {\em et al.}\cite{gan}  \\
PSR~J1614-2230 &0.0062&$9.69\pm0.2$&$1.97\pm0.04$&9.7& 1.97&Gangopadhyay {\em et al.}\cite{gan} \\
Vela~X-1 &0.0053&$9.56\pm0.08$&$1.77\pm0.08$&9.56 & 1.77&Gangopadhyay {\em et al.}\cite{gan} \\
Cen-X3 &0.0046&$9.178\pm0.13$&$1.49\pm0.08$&9.18&1.5&Gangopadhyay {\em et al.}\cite{gan}\\
\hline
\end{tabular*}
\end{table*}

\begin{table*}
\caption{Values of physical parameters for different compact star models. }
\begin{tabular*}{\textwidth}{@{\extracolsep{\fill}}llllll@{}}
\hline
Compact Star & central density & surface density & $u$ & $z_s$\\
& gm~cm$^{-3}$ & gm~cm$^{-3}$\\ \hline
4U~1538-52  &$1.89\times10^{15}$&$0.72\times10^{15}$&0.168&0.228  \\
PSR~J1614-2230 &$1.99\times10^{15}$&$0.68\times10^{15}$&0.299&0.579  \\
Vela~X-1 &$1.71\times 10^{15}$&$0.67\times 10^{15}$& 0.273 &0.484\\
Cen-X3&$1.49\times10^{15}$ &$0.67\times 10^{15}$& 0.241  & 0.389     \\ \hline
\end{tabular*}
\end{table*}

\subsection{\emph{Stability Analysis}}
Let us now address the issue for stability analysis of the solution.
In order to study the dynamical stability of the transition layer of these dark energy stars
we consider a linear perturbation around those static solutions. It is interesting to note that Eq. (23),can be altered to obtain an equation of motion of the thin shell, as follows,
\begin{equation}
\dot{R}^2+V(R)=0,
\end{equation}
with V (a) given by
\begin{equation}
V(R)=1-\frac{m(R)+M}{R}-\left(\frac{m_s(R)}{2R}\right)^2-\left(\frac{M-m(R)}{m_s(R)}\right)^2.
\end{equation}
Note that the potential function V(a) helps us to determine the stability region for
the thin shell under our linear perturbation. Now we shall consider the
Taylor series expansion around the static solution $R_0$, upto second order, we obtain
\begin{eqnarray}
V(R)& = & V(R_0)+(R-R_0)V^\prime(R_0)+\frac{(R-R_0)^2}{2}V^{\prime\prime}(R_0)+\nonumber\\
&\;& +\mathcal{O}[(R-R_0)^3],\nonumber\\
\end{eqnarray}
where prime corresponding to a derivative with respect to R. According to the standard method
we are linearizing around the static radius $R = R_0$, we must have $V(R_0)= 0$ and  $V^\prime(R_0)=0$.
It is straightforward to see that $V^\prime(R_0)=0$ gives the relation

\begin{eqnarray}
\left(\frac{m_s(R_0)}{2R_0}\right)^\prime= \Phi &= &\frac{R_0}{m_s(R_0)}
\left[~-\left(\frac{m(R_0)+M}{R_0}\right)^\prime\right.\nonumber\\
&\;&\left.-2\left(\frac{M-m(R_0)}{m_s(R_0)}\right)\left(\frac{M-m(R_0)}{m_s(R_0)}\right)^\prime~\right].\nonumber\\
\end{eqnarray}
With this definition the second derivative $V^{\prime\prime}(R_0)$ can be written as
\begin{eqnarray}
V^{\prime\prime}&=&-\left(\frac{m+M}{R_0}\right)^{\prime\prime}-2\left[\left(\frac{m_s}{2R_0}\right)^\prime\right]^2
-2\left(\frac{m_s}{2R_0}\right)\left(\frac{m_s}{2R_0}\right)^{\prime\prime}\nonumber\\
&&-2\left[\left(\frac{M-m}{m_s}\right)^\prime~\right]^2-2\left(\frac{M-m}{m_s}\right)
\left(\frac{M-m}{m_s}\right)^{\prime\prime}.
\end{eqnarray}
Thus, a static configuration demands that the first two terms of the Taylor series expansion
vanish, and the first non-zero term in the expansion for equation of motion of the
thin shell may be written as (considering the Eq. (41))
\begin{eqnarray}
\dot{R}^2+\frac{1}{2}V^{\prime\prime}(R_0)\left(R-R_0\right)^2+\mathcal{O}[(R-R_0)^3] = 0,
\end{eqnarray}
To ensure the stability of static configuration at R = $R_0$, the second derivative of the
potential function must be positive, i. e., $V^{\prime\prime}(R_0) > 0$. For the sake of
simplicity we rearrange the Eq. (45), which turns out
\begin{equation}
\Rightarrow~ V^{\prime\prime} = \Pi-2\Phi^2-\frac{m_s}{R_0}\left( \Upsilon-4\pi~\sigma^\prime\eta \right) \Big|_{R_0}.
\end{equation}
In this regard, we proceed with notational simplicity, which turns out to be
\begin{equation}
\left(\frac{m_s}{2R}\right)^{\prime\prime} = \Upsilon-4\pi~\sigma^\prime\eta,
\end{equation}
with
\begin{eqnarray}
\Pi & = & -\left(\frac{m+M}{R_0}\right)^{\prime\prime}-2\left[\left(\frac{M-m}{m_s}\right)^\prime\right]^2+\nonumber\\
&\;& -2\left(\frac{M-m}{m_s}\right)\left(\frac{M-m}{m_s}\right)^{\prime\prime}.\nonumber\\
\end{eqnarray}

We focus our attention on small velocity perturbations, thus assuming $\eta(R_0)=\eta_0$
and using Eq.(47) for $V^{\prime\prime}(R_0)> 0$ we have
\begin{equation}
\eta_0\frac{d\sigma^2}{dR}\Big|_{R_0}>\frac{\sigma}{2\pi}\left[\Upsilon+\frac{R_0}{m_s}(2 \Phi^2-\Pi)\right] = \Gamma ~(\mathrm{say}).
\end{equation}
Let us begin by examining the stable equilibrium regions by rewriting the Eq. (50), in the
suggestive form as follows :
\begin{equation}
\eta_0>\Gamma\left(\frac{d\sigma^2}{dR}\Big|_{R_0}\right)^{-1}, ~~~ \mathrm{if}~~~ \frac{d\sigma^2}{dR}\Big|_{R_0} > 0,
\end{equation}
\begin{equation}\label{eq:less}
\eta_0<\Gamma\left(\frac{d\sigma^2}{dR}\Big|_{R_0}\right)^{-1}, ~~~ \mathrm{if}~~~ \frac{d\sigma^2}{dR}\Big|_{R_0} < 0,
\end{equation}
Let us make a comment concerning the dark energy stars by choosing
specific mass functions, and deduce the stability region by
considering the inequalities given in Eqs. (51-52). As it
was pointed out in Ref. \cite{lobo, pb15}, we can find the stability regions
by the qualitative plots. The result is illustrated in Fig. \textbf{7}, which indicates that the
stability region is given below the surface by Eq. (52).

\section{Results and Discussion}

In our present paper we have challenged and provided an alternative to the idea of black holes in the form of the notion of dark energy star.Though this is a strong statement to make, considering the fact that the concept of black hole and its event horizon is almost universally accepted, yet any kind of scientific theory must be strengthened by subjecting it to careful scrutiny and cross verification with the help of alternate ideas.In spite of numerous good research favouring Black hole horizons, they still introduce a number of theoretical problems which have not yet been satisfactorily resolved. Hence it has been argued by several authors that after the gravitational collapse of massive stars, different objects could be formed other than black holes. One proposal, which was initiated recently, to dissolve the
singularity and horizon problem is the gravastar (gravitational vacuum star ) model
proposed by Mazur and Mottola \cite{mazur}, which has an effective phase transition at/near where the
event horizon is expected to form. It has recently been shown that one could ideally
generalize the gravastar picture by matching an interior solution corresponding a
dark energy equation of state, to an exterior Schwarzschild metric at a junction interface.
Such a spherically symmetric model should not possess a horizon and it should possess a
static equilibrium solution, sometimes referred to as a dark energy star
(see for instance Ref. \cite{lobo} for review).

We have studied the model describing compact gravitating configurations
governed by the dark energy equation of state. We have analyzed the stellar
configuration by imposing a specific choice of mass function, which
is uniform in the stellar interior. The solutions set thus obtained are
correlated with a set of astrophysical objects in connection to direct
comparison of some strange/compact star candidates like X-ray pulsar Cen-X3,
X-ray burster 4U 1538-52, and Xray sources 4U 1538-52. In order to compare
the dark energy star with observational data, we specially considered the X-ray pulsar
Vela X - 1, whose estimated mass M = $1.77 M_{\odot}$ and radius $9.56$.
On substituting these values into the relevant equations
we obtained the expression for energy density, radial and transverse pressures, gravity profile
and anisotropic parameter for the model, where the energy density
is maximum at the center and gradually decreases away towards the boundary (see Fig. (1)).
All this results have been shown in more instructive way by graphical
representation shown in Figs. 1-7.

Furthermore, we emphasize the results in more details, with observational data
by Gangopadhyay {\em et al.} \cite{gan} of some well known pulsars like
4U 1538-52, PSR J1614-2230, Vela X - 1 and Cen X - 3.
For this purpose, we produce data sheet for the purpose of comparison between
present model stars and the known compact objects in Table~I and ~II.
It is to be noted that we have set $G = c = 1$, while solving Einstein's
equations as well as for plotting all the figures. As one can see, the results
extracted in this theory is very much compatible with
the results obtained through the observations and the obtained ratio of mass-radius
for different strange stars that lies in the proposed range by Buchdahl \cite{buchdahl}.
Next, we examined the surface redshift ($Z_s$) of the different compact stars are
of finite values and vanishes out side of the star (see Table - II), which lies
in the proposed range by Barraco \& Hamity \cite{hamity}, and hence are physically acceptable
with those of observations. We conclude that it is possible to obtain
the existence of dark energy stars, however, it is important to understand the nature and general properties
of compact objects by thorough fine tuning.

\subsection*{Acknowledgments}
FR and AB would like to thank the authorities of the Inter-University Centre for Astronomy
and Astrophysics, Pune, India for providing the research facilities. FR is also
thankful to DST-SERB for financial support.

\end{document}